\documentclass[preprint,12pt]{elsarticle}

\usepackage{graphicx}
\usepackage{amssymb}

\usepackage[table]{xcolor}
\usepackage[colorinlistoftodos]{todonotes}
\usepackage[caption=false,font=footnotesize]{subfig}

\usepackage{lineno}

\journal{Nuclear Instrumentation Methods}

\begin{document}

\begin{frontmatter}


\title{Neutron irradiation effect on SiPMs up to $\Phi_{neq}$~=~5$\times$10$^{14}$~cm$^{-2}$}



\author[mysecondaryaddress]{M. Centis Vignali}
\author[mymainaddress]{E. Garutti\corref{mycorrespondingauthor}}
\cortext[mycorrespondingauthor]{Corresponding author} \ead{Erika.Garutti@physik.uni-hamburg.de}
\author[mymainaddress]{R. Klanner}
\author[mymainaddress]{D. Lomidze}
\author[mymainaddress]{J. Schwandt}
\address[mymainaddress]{Hamburg University, Luruper Chaussee 149, 22761 Hamburg, Germany}
\address[mysecondaryaddress]{CERN, Geneva, Switzerland}

\begin{abstract}
Silicon Photo-Multipliers (SiPM) are becoming the photo-detector of choice for increasingly more particle detection applications, from fundamental phys-ics, to medical and societal applications. One major consideration for their use at high-luminosity colliders is the radiation damage induced by hadrons, which leads
to a dramatic increase of the dark count rate.  KETEK SiPMs have been
exposed to various fluences of reactor neutrons up to $\Phi_{neq}$~=~5$\times$10$^{14}$ \rm{cm}$^{-2}$ (1~\rm{MeV} equivalent
neutrons). 
Results from the I-V, and C-V measurements for temperatures between \rm{$-$30$^\circ$C} and \rm{$+$30$^\circ$C} are presented. We propose a new method to quantify the effect of radiation damage on the SiPM performance. Using the measured dark current the single pixel occupation probability as a function of temperature and excess voltage is determined. From the pixel occupation probability the operating conditions for given requirements can be optimized. The method is qualitatively verified using current measurements with the SiPM illuminated by blue LED light.  

\end{abstract}

\begin{keyword}
SiPM \sep Radiation damage \sep Neutron irradiation


\end{keyword}

\end{frontmatter}


\section{Introduction}
\label{S:1}
Radiation damage of silicon by hadrons has been extensively
studied for electronics and sensors \cite{Srour:2013kg,Lindstrom:2003il}. In the Si bulk,
defect states are formed, which change the effective doping,
reduce the carrier mobilities and lifetimes, and increase
the generation rate. The increased generation rate causes an increase
in dark-count-rate ($DCR$), which is the biggest limitation
for use of SiPMs in a high radiation environment. To better quantify the effect of the $DCR$ increase on the SiPM performance, the pixel occupation probability is determined using the measured dark current and the values of the quenching resistance. The pixel occupation probability is defined as the the probability that a Geiger discharge occurs in a pixel in a given time interval. From the pixel occupation probability the decrease of the dynamic range of the SiPM due to dark counts can be determined. The results are compared to SiPM current measurements with the SiPM illuminated by a blue LED. \\
In this paper current-voltage measurements of SiPMs irradiated with neutrons to fluences between 0 and $\Phi_{neq}$~=~5$\times$10$^{13}$~cm$^{-2}$, with and without illumination by a blue LED, and temperatures between \rm{$-$30$^\circ$C} and \rm{$+$30$^\circ$C} are presented. From these data the temperature and fluence dependence of characteristic SiPM parameters, like breakdown voltage, pixel occupancy, and reduction of the photo-detection efficiency are determined. 

\section{SiPMs, irradiation and measurements}
\label{S:2}

The SiPMs investigated were fabricated by KETEK \cite{Ketek}. They
consist of 4384 pixels of 15~$\times$~15~$\mu$m$^2$, a breakdown voltage
of about 27.5~V, a depth of the amplification region,
as determined by capacitance-voltage (C-V) measurements,
of  $<$~1~$\mu$m, and a poly-silicon quenching resistance of
 550~k$\Omega$ with a sample-to-sample spread of $\pm$30\%. For more details see \cite{Chmill:2016msk}. The neutron irradiations
were performed at room temperature without applied bias
at the TRIGA Research Reactor of the JSI, Ljubljana. The samples were transported cold to Hamburg after irradiation and stored in a refrigerator at \rm{$-$30$^\circ$C}. No annealing was applied to the samples before measurement. However, each measurement cycle took approximately 2 hours at one given temperature, and in particular the measurements at \rm{$+$30$^\circ$C} cause annealing. The sample are kept in the refrigerator when not being measured. \\
The following measurements
were performed on a temperature-controlled chuck in a dry atmosphere: current-voltage (I-V) for
forward and reverse voltages; temperatures between \rm{$-$30$^\circ$C}
and \rm{$+$30$^\circ$C} without and with illumination by LED light of 470 nm. 
\begin{figure}[h]
\centering
\subfloat[]{
\includegraphics[width=0.45\linewidth]{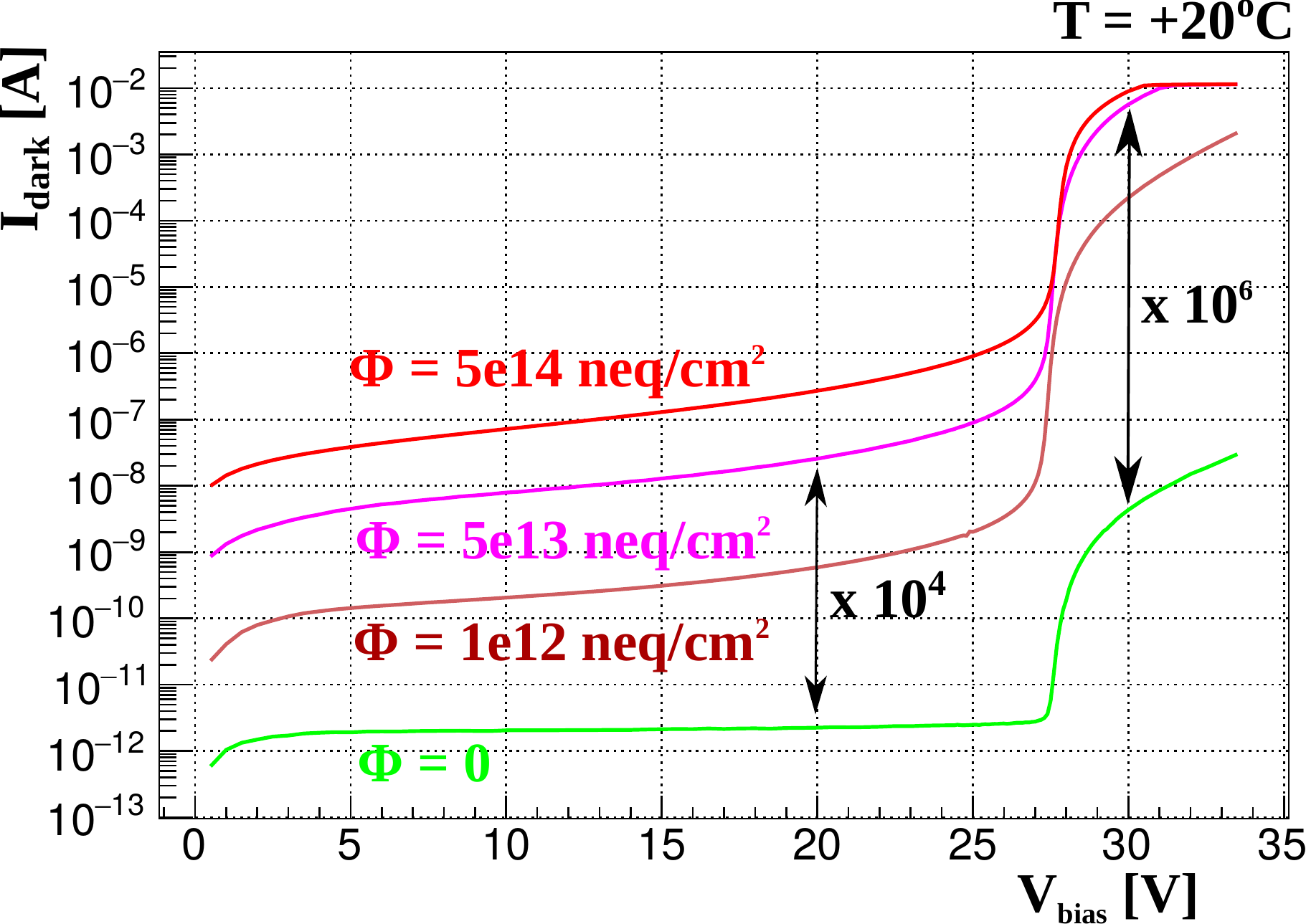}}
\subfloat[]{
\includegraphics[width=0.45\linewidth]{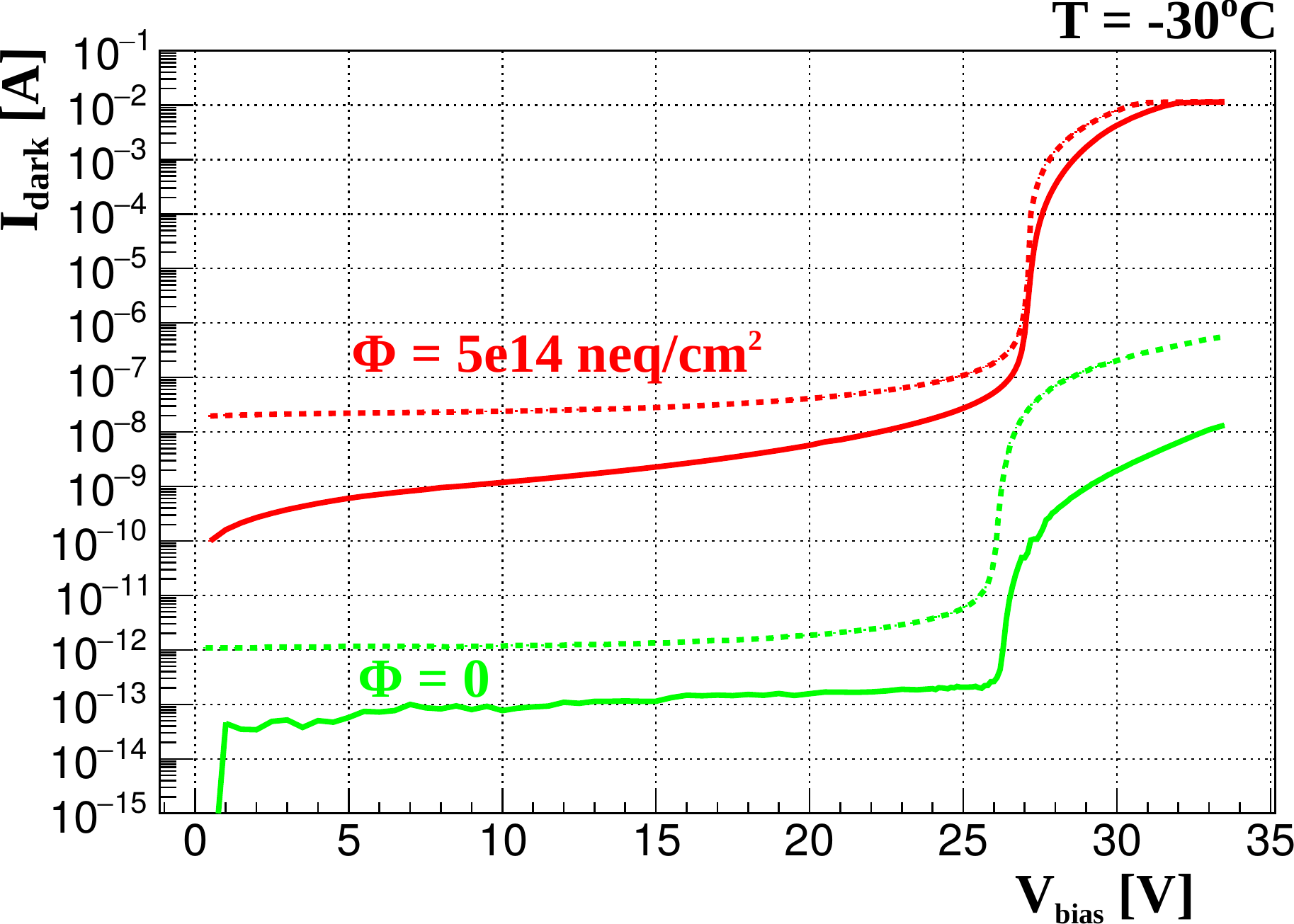}}
\caption{Measured I-V curves at various fluences, (a) at \rm{$+$20$^\circ$C}, in the dark, (b) at \rm{$-$30$^\circ$C} in the dark (solid lines) and with LED illumination (dashed lines). The related fluences are reported next to the curves.}
\label{fig:IV}
\end{figure}

\section{Breakdown voltage}
\label{S:vbd}
\begin{figure}[h]
\centering
\subfloat[]{
\includegraphics[width=0.43\linewidth]{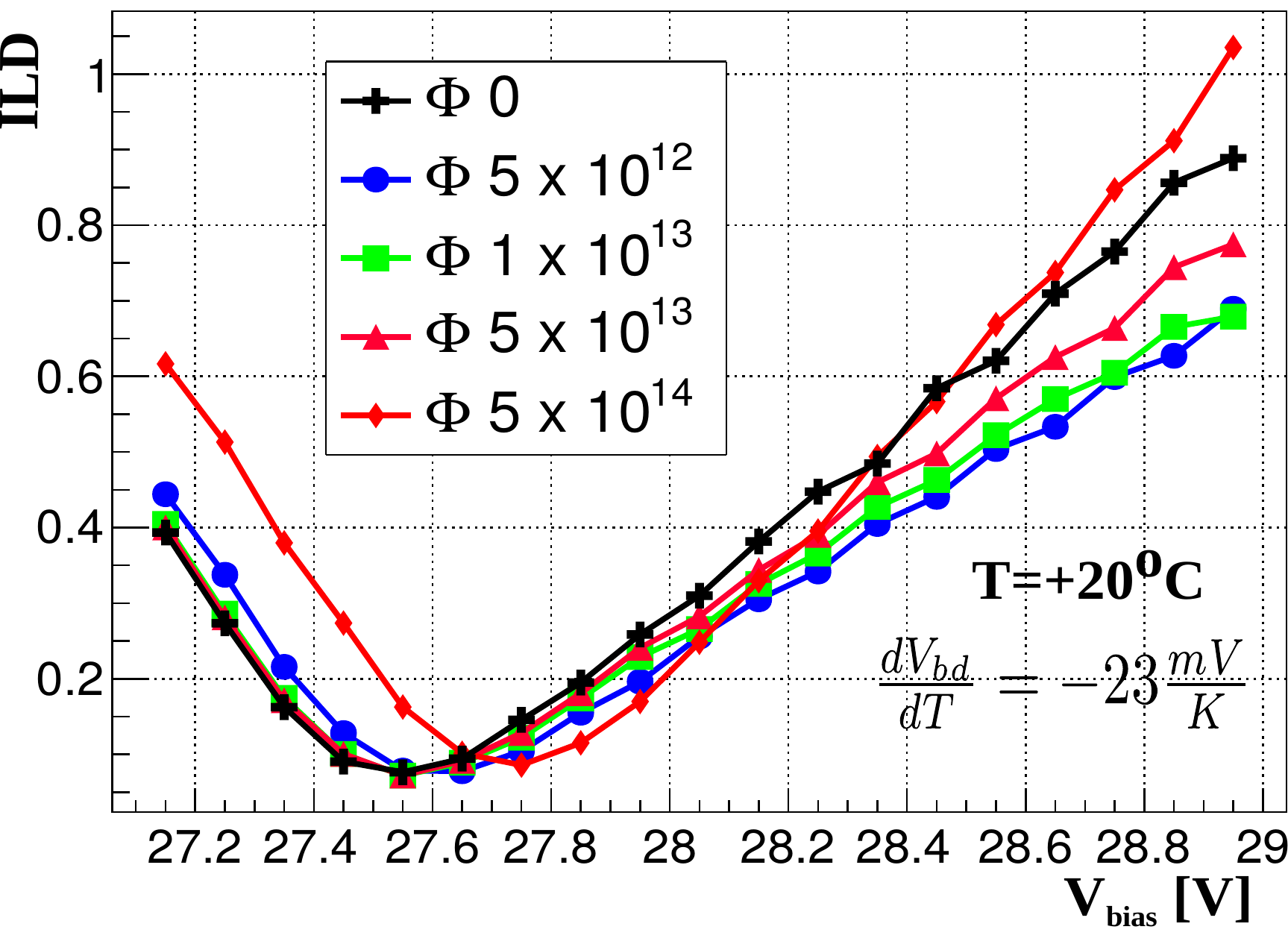}}
\subfloat[]{
\includegraphics[width=0.47\linewidth]{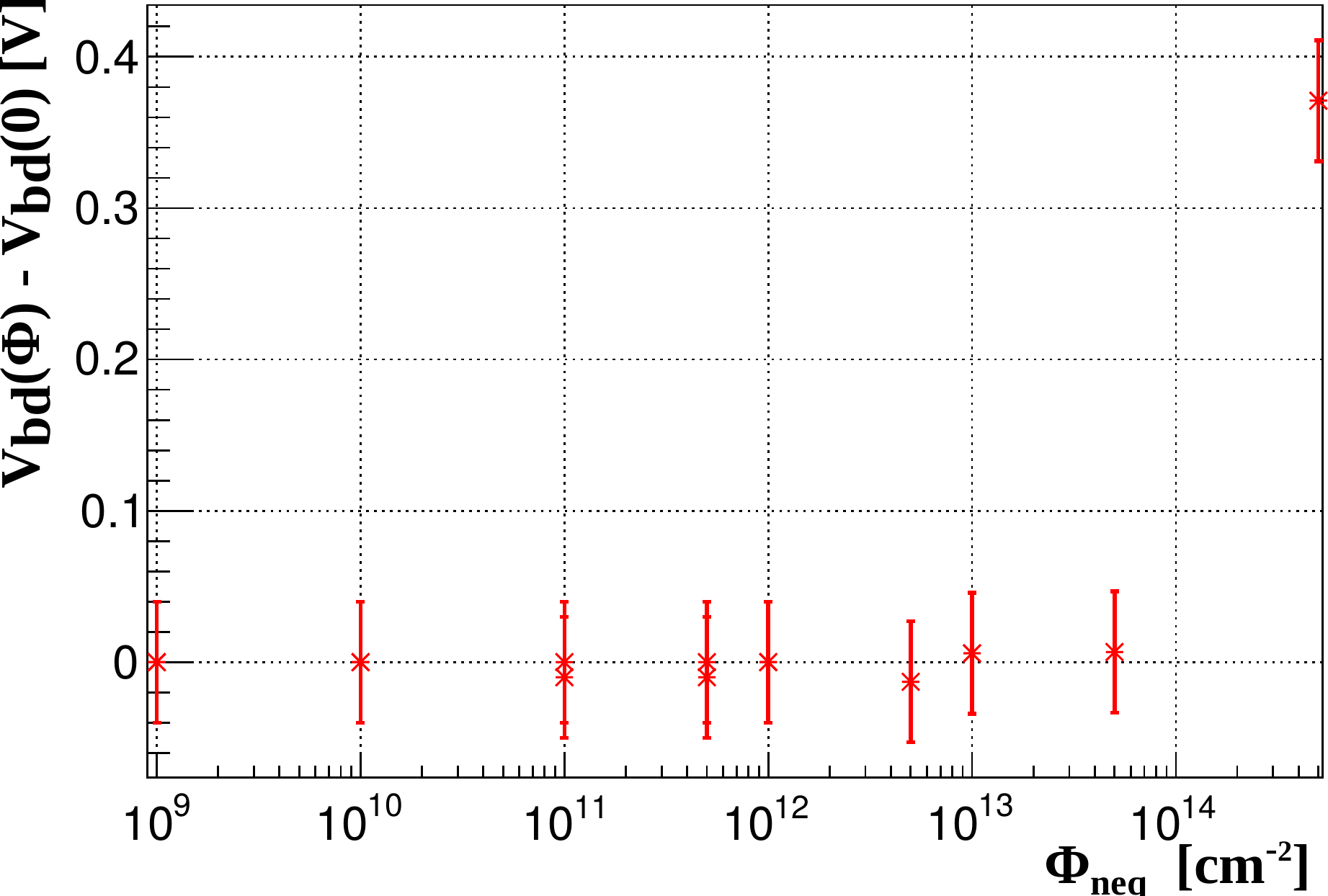}}
\caption{a) ILD curves computed from the I-V curves of Fig.~\ref{fig:IV} (a). b) difference of $V_{bd}$ after and before neutron irradiation as function of fluence.} 
\label{fig:ILD}
\end{figure}

The breakdown voltage $V_{bd}$ for each set of measurement is determined using the method of the minimum of the inverse logarithmic derivative (ILD) as discussed in \cite{Chmill:2016msk}. Fig.~\ref{fig:ILD} (a) presents ILD for the I-V curves shown in Fig.~\ref{fig:IV} (a). 
The difference of $V_{bd}$ after and before neutron irradiation as function of irradiation fluence are presented in Fig.~\ref{fig:ILD} (b) for T~=~\rm{$+$20$^\circ$C}. Up to $\Phi_{neq}$~=~5$\times$10$^{13}$~cm$^{-2}$, no change is observed in the value of $V_{bd}$ within the uncertainty of about 40~mV. The value of $V_{bd}$ after irradiation with $\Phi_{neq}$~=~5$\times$10$^{14}$~cm$^{-2}$ neutrons is higher by about 350~\rm{mV} compared to the non-irradiated SiPM. Note that in \cite{Chmill:2016msk} we reported a difference of about 1~V for the $V_{bd}$ determined from I-V curves and the $V_{bd}^G$ determined from  gain ($G$) vs. voltage curves, for this specific SiPM. Therefore the results from of Fig.~\ref{fig:ILD} (b) do not allow to draw conclusions on the fluence dependence of the SiPM gain, which is expected to be proportional to $V-V_{bd}^G$. Further work is required to establish a method to determine  $V_{bd}^G$ for irradiated SiPMs.

\section{Photo-detection}
\label{S:PDE}
\begin{figure}[h]
\centering
\subfloat[]{
\includegraphics[width=0.5\linewidth]{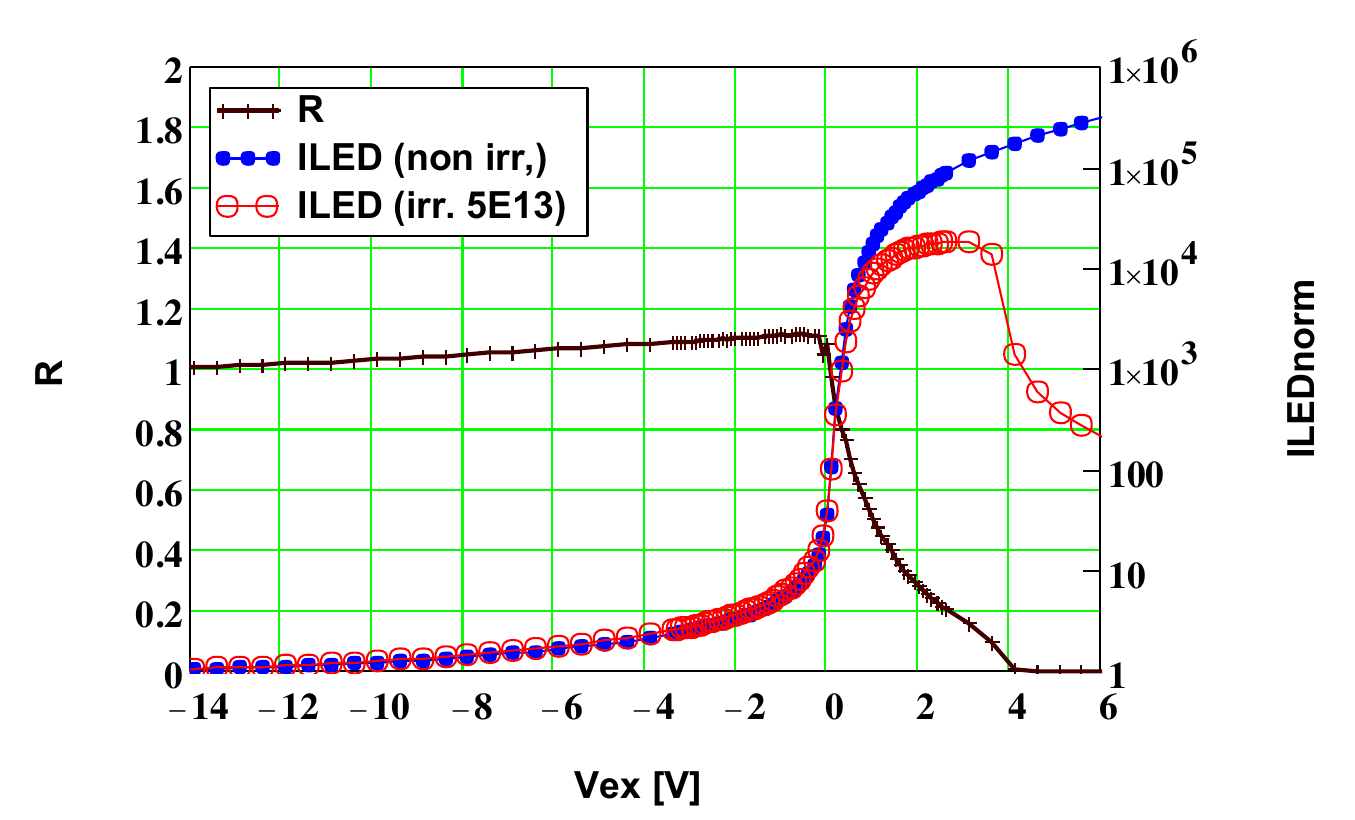}}
\subfloat[]{
\includegraphics[width=0.5\linewidth]{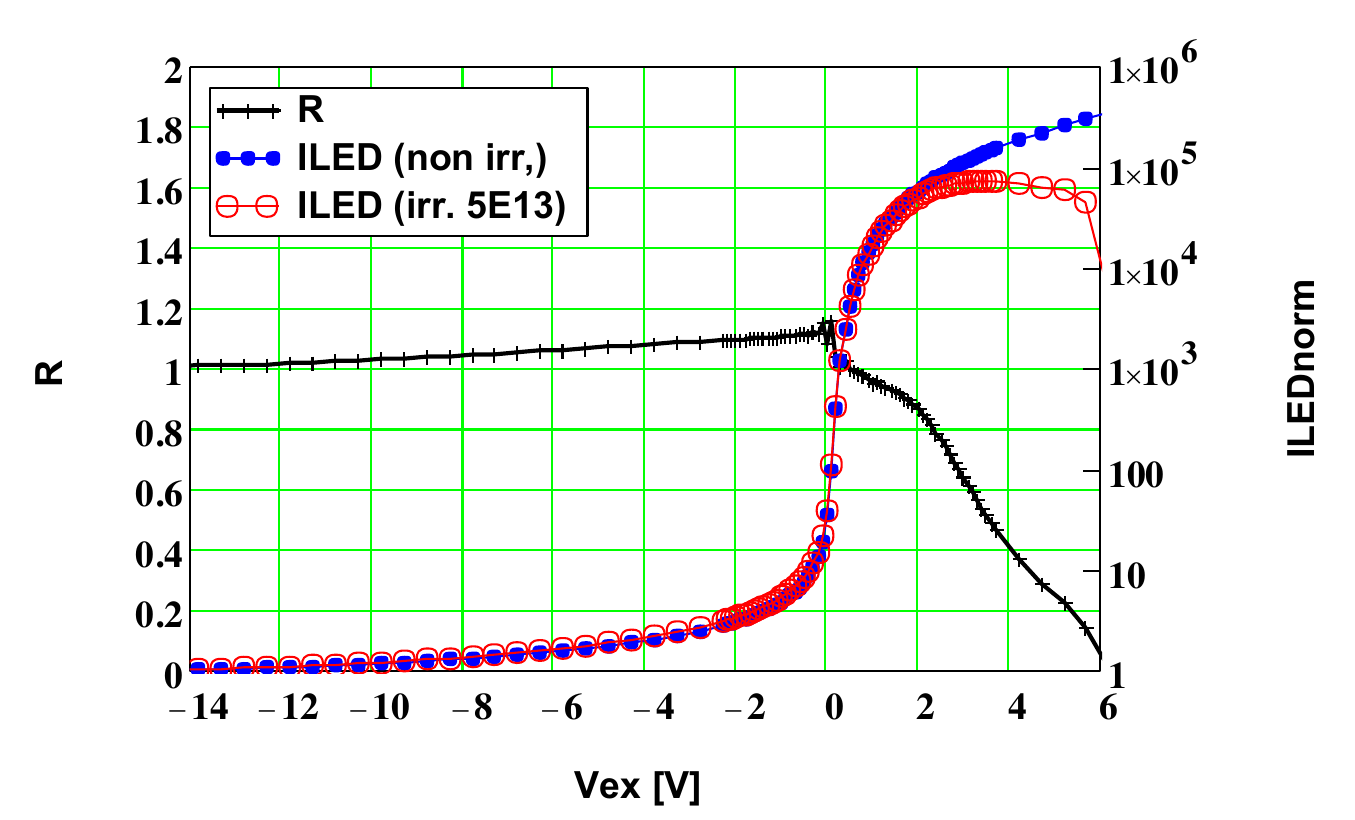}}
\caption{$I_{LED}^{norm}$ for the non-irradiated  SiPM (solid dots) and the SiPM irradiated to $\Phi_{neq}$~=~5$\times$10$^{13}$~cm$^{-2}$ (open circles) with the y-scale on the right, and their ratio, R, with the scale on the left. (a) at \rm{$+$20$^\circ$C} and (b) at \rm{$-$30$^\circ$C}. For details see text.}
\label{fig:PDE}
\end{figure}

One relevant question for SiPM applications is, how does the photo-detection change as a function of fluence and temperature either because of  changes in the SiPM electronic parameters (signal duration, $\tau$, electric field, $V_{bd}$, $PDE$, correlated noise ($CN$)) or because of the increase in $DCR$. Fig.~\ref{fig:PDE} presents the normalized photo-current 

\begin{equation}
\label{eq:Inorm}
I_{LED}^{norm}=\frac{I_{LED}^{meas}(V_{bias})-I_{dark}(V_{bias})}{I_{LED}^{meas}(V_{bias}=10~V)-I_{dark}(V_{bias}=10~V)}
\end{equation}

\noindent of the SiPM before and after irradiation to $\Phi_{neq}$~=~5$\times$10$^{13}$~cm$^{-2}$. At $V_{bias}=10~V$ the SiPM gain is assumed to be 1. If the additional pixel occupancy by the LED photons is ignored $I_{LED}^{norm}$ can be related to SiPM parameters by $I_{LED}^{norm} \approx A^*_{prob} \cdot G \cdot (1+CN) $, with $A^*_{prob}$ being the Geiger discharge probability multiplied by the probability that the corresponding pixel is not occupied by a Geiger discharge. The ratio 
\begin{equation}
\label{eq:R}
R = \frac{I_{LED}^{norm}(\Phi_{neq})}{I_{LED}^{norm}(\Phi_{neq}=0)} 
\end{equation}
\noindent should be equal to 1 if the product $A^*_{prob} \cdot G \cdot (1+CN) $ is not affected by the irradiation. While this was the case within $<$~10\% up to $\Phi_{neq}$~=~10$^{12}$~cm$^{-2}$, as demonstrated in~\cite{IEEE:2016mcv}, it is not anymore true for $\Phi_{neq}$~=~5$\times$10$^{13}$~cm$^{-2}$. Above breakdown $R$ drops quickly to zero, indicating a rapid decrease in effective photo-detection efficiency. Cooling to \rm{$-$30$^\circ$C} increases the excess voltage range for a given lower limit of $R$. To understand the cause of this signal loss the pixel occupation probability is investigated.

\section{Pixel occupation probability}
\label{S:pix}
We introduce the pixel occupation probability due to dark counts $\eta_{DC}$, as the probability of a Geiger discharge in a pixel in a time interval $\Delta t$. It is related to the $DCR$ by  
\begin{equation}
\label{eq:DCR}
DCR \cdot (1+CN) \approx \frac{N_{pix}}{\tau}\cdot\eta_{DC}, 
\end{equation}
where we have taken  $\Delta t = \tau = R_q \cdot C_{pix}$ the recovery time of the SiPM pulse. 
One can express the measured $I_{dark}$ in terms of the pixel occupation probability: 
\begin{equation}
\label{eq:Idark}
I_{dark} = q_0 \cdot G \cdot DCR \cdot (1+CN) \approx q_0 \cdot \frac{C_{pix}V_{ex}}{q_0} \cdot \frac{N_{pix}}{\tau}\cdot \eta_{DC}. 
\end{equation}
Eq.~\ref{eq:Idark} can be rewritten as:
\begin{equation}
\label{eq:Idark2}
\eta_{DC} = \frac{I_{dark}}{V_{ex}} \cdot \frac{R_q}{N_{pix}}.
\end{equation}
We note that all quantities in Eq.~\ref{eq:Idark2}, $I_{dark}$, $R_q$ and $V_{bd}$ can be determined from I-V measurements for forward and reverse bias. In particular, $R_q$ is taken from $\frac{dI}{dV}$ calculated at the highest forward bias voltage below current limit of 1.7~\rm{V}. \\
We also note that for $\eta_{DC} \rightarrow 1$  the voltage drop due to  $I_{dark}$ over $R_q/N_{pix} = V_{ex}$; thus the pixel voltage never recovers, i.e. $V_{bias}-V_{ex} = V_{bd}$. 

Assuming Poisson statistics, we calculate $\mu_{DC}= -ln \left(1-\eta_{DC}\right)$, the average number of $e-h$ pairs which in a time interval $\tau$ would produce a Geiger discharge in a pixel not already occupied by a discharge. The quantity $\mu_{DC}/\tau$ is directly related to the charge carrier generation rate (and thus can be simulated using Shockley-Read-Hall statistics including field enhancement).


\begin{figure}[h]
\centering
\subfloat[]{
\includegraphics[width=0.5\linewidth]{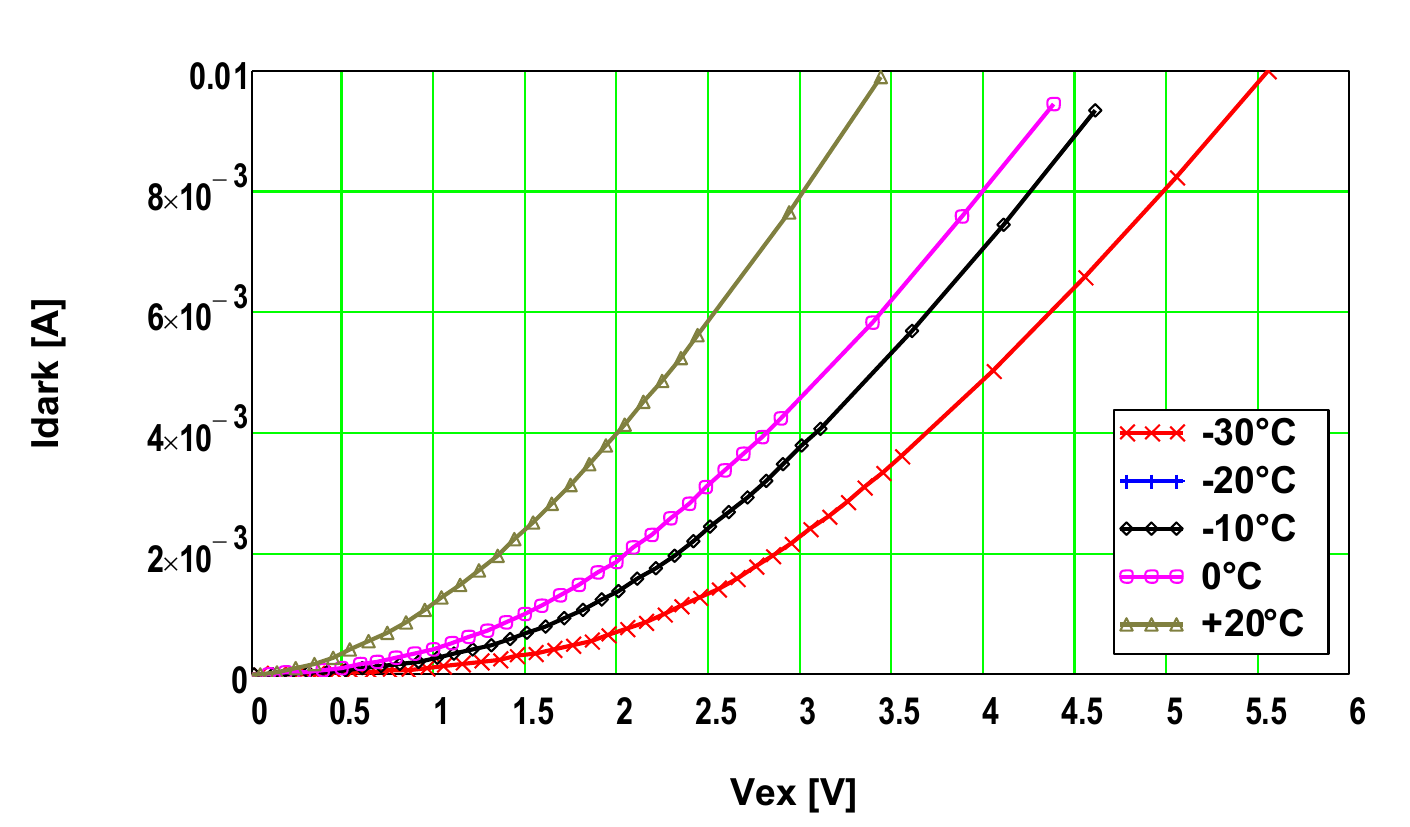}}
\subfloat[]{
\includegraphics[width=0.47\linewidth]{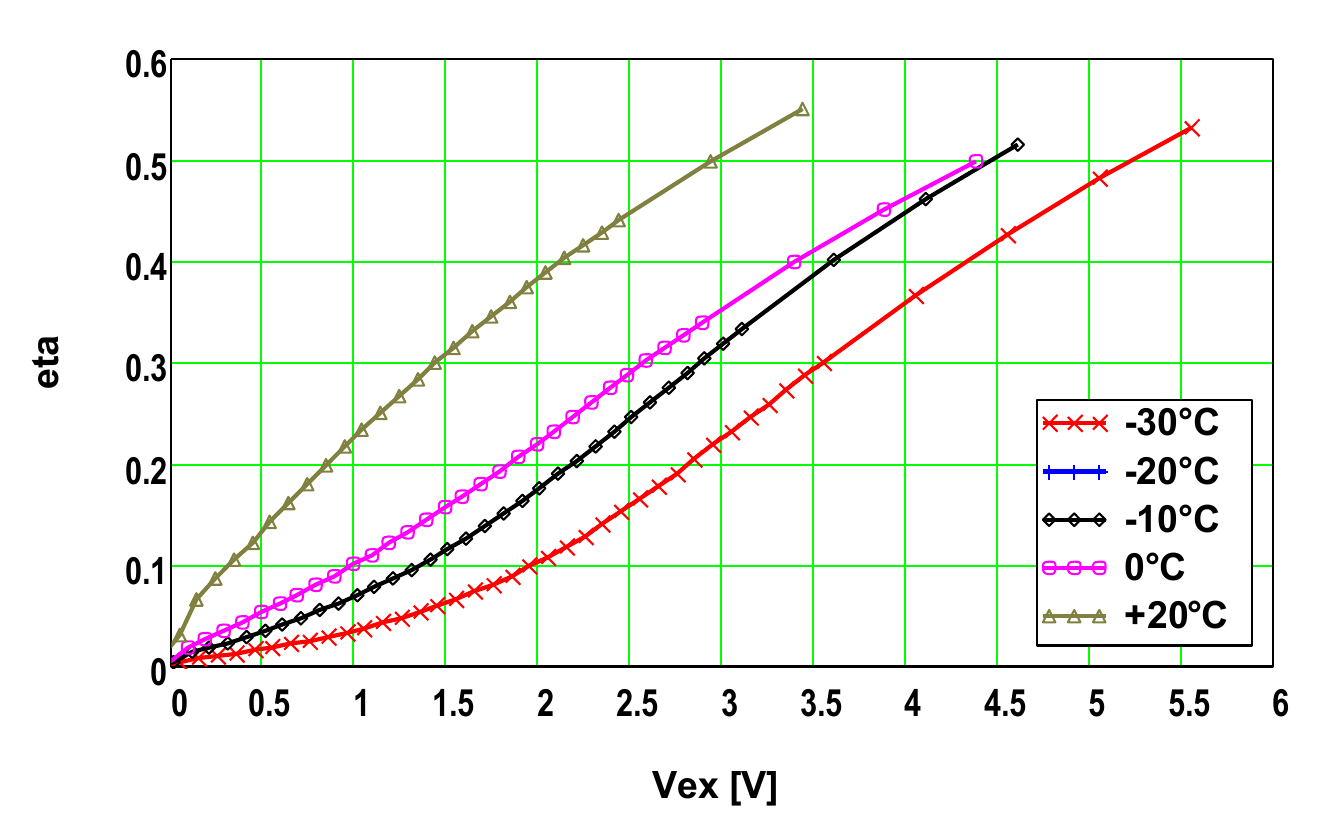}}
\caption{a) $I_{dark}$ versus excess bias voltage for a SiPM irradiated to $\Phi_{neq}$ = 5$\times$10$^{13}$ cm$^{-2}$. b) Pixel occupation probability $\eta_{DC}$ for various temperatures from +20~$^o$C to \rm{$-$30$^\circ$C}.}
\label{fig:mu}
\end{figure}

Fig.~\ref{fig:mu}a shows the measured $I_{dark}$ versus excess bias voltage for a SiPM irradiated to $\Phi_{neq}$ = 5$\times$10$^{13}$ cm$^{-2}$ for various temperatures from \rm{$+$20$^\circ$C} to \rm{$-$30$^\circ$C}. Fig.~\ref{fig:mu}b presents the results for the pixel occupation probability $\eta_{DC}$ extracted from Fig.~\ref{fig:mu}a. The curves indicate that for instance for an operation temperature of \rm{$-$30$^\circ$C}, this particular SiPM has a pixel occupation probability of 20\% if operated at 2.5 V excess bias. We conclude that the increase of the pixel occupation probability, $\eta_{DC}$, with temperature and fluence is responsible for the rapid decrease of the photo-detection efficiency, as function of fluence and temperature, as demonstrated by the decrease of $R$ shown in Fig.~\ref{fig:PDE}.

\section{Conclusions}
Different characteristics of KETEK SiPMs irradiated with
neutrons up to a fluence of $\Phi_{neq}$ = 5$\times$10$^{14}$~cm$^{-2}$ were extracted from current-voltage measurements  with and without illumination with a blue LED and temperatures between \rm{$-$30$^\circ$C} and \rm{$+$30$^\circ$C}.
The values of the breakdown voltage is not changed up to $\Phi_{neq}$ = 5$\times$10$^{13}$~cm$^{-2}$, whereas an increase of $V_{bd}$ is observed for $\Phi_{neq}$ = 5$\times$10$^{14}$~cm$^{-2}$.
For high neutron fluences, the $DCR$ by far exceeds the values for which the standard methods of $DCR$ determination using pulse-height spectra can be applied. Therefore the method of pixel occupation probability is introduced, which allows to characterize the reduction of photo-detection efficiency due to high DCRs.  As an example, the specific KETEK SiPM irradiated to $\Phi_{neq}$ = 5$\times$10$^{13}$~cm$^{-2}$ has a pixel occupation probability of 20\% if operated at $V_{ex}$~=~2.5~\rm{V} at \rm{$-$30$^\circ$C}, and thus can be used as photo-detector, however with a significantly reduced dynamic range compared to the non-irradiated SiPM.

\section{Acknowledgements}
The authors would like to thank Florian Wiest and his colleagues from KETEK for providing the SiPMs samples and for the fruitful discussions. This project has received funding for the samples irradiation from the European Union's Horizon 2020 Research and Innovation programme under Grant Agreement no.\ 654168.





\bibliographystyle{elsarticle-num-names}
\bibliography{elsarticle-template-1-num.bib}







\end{document}